\documentstyle[11pt,aaspp4,epsf]{article}

\begin{document}

\def\ni{\noindent}
\def\t{\thinspace}
\def\i{\indent$\bullet$\ }
\def\b{$\bullet$\ }
\def \rns {$R_{\rm NS}$}
\def \egret {EGRET\ }
\def \al {$\alpha$}
\def \ze {$\zeta$}
\def  \delob {$\Delta^{\rm peak}_{\rm obs}$}
\def  \del {$\Delta^{\rm peak}$}
\def  \thb {$\theta_b$}
\def  \eps {$\epsilon$}
\def  \hi  {$h_{\rm init}$}
\def  \et  {$\epsilon_{\rm turn}$}
\def  \ee  {$\epsilon_{\rm esc}$}

\def\egret{{\it EGRET~}}
\def\rosat{{\it ROSAT~}}
\def\cosb{{\it COS-B~}}
\def\vela{{PSR0833-45~}}
\def\TeV{~\rm{TeV}}
\def\GeV{~\rm{GeV}}
\def\MeV{~\rm{MeV}}
\def\keV{~\rm{keV}}
\def\eV{~\rm{eV}}
\def\K{~\rm{K}}
\def\G{~\rm{G}}
\def\s{~\rm{s}}
\def\ms{~\rm{ms}}
\def\cm{~\rm{cm}}
\def\Lg{$L_\gamma$}
\def\Lx{$L_X$}
\def\edot{$L_{\rm sd}$}

\title{The Question of the Peak Separation in the Vela Pulsar\footnote{to be published in
Proc. 19th Texas Symposium (Mini-symposium: ``Pulsars and Neutron Stars")}}

\author{J. Dyks$^1$, B. Rudak$^1$ and T. Bulik$^2$\\Nicolaus
Copernicus Astronomical Center\\ $^1$ Rabia{\'n}ska 8, 87100 Toru{\'n}, Poland
\\ $^2$ Bartycka 18, 00716
Warsaw, Poland} 

\begin{abstract}
We calculate gamma-ray spectra of pulsars for a set of electron acceleration models
within the polar-cap scenario. We discuss our results in the light of the 
recent analysis (Kanbach 1999) of temporal properties of the Vela pulsar
in the EGRET energy range.

\end{abstract}

\section{Introduction}

In a recent analysis of gamma-ray double-peak pulses of the Crab, the Vela and Geminga,
as detected by \egret, Kanbach (1999) 
pointed to an intriguing possibility that for the Vela pulsar
the phase separation ($\Delta^{\rm peak}$) between the two peaks
may actually be energy dependent. The energy-averaged value of the separation is very large in
all three objects, between 0.4 and 0.5 (Fierro, Michelson \& Nolan 1998), 
and the effect by itself is of the order of
a few percent or less. In the case of the Vela pulsar (the object with
the best photon statistics), the plot 
of $\Delta^{\rm peak}$ against energy (Fig.2 middle panel of Kanbach 1999) shows that
$\Delta^{\rm peak}$ is decreasing by about $5\%$ 
over 20 energy intervals covering the range between $\sim 50\MeV$ and $\sim 9\GeV$. 
The scatter of the points seems to us, however, also
consistent with the separation staying at a constant level of $0.43$, 
provided that we reject the lowest and the highest energy interval. 

The problem raised by Kanbach is important by itself
from a theoretical point of view, regardless of whether his finding becomes
well established or not. With future high-sensitivity missions like GLAST,
any firm empirical relation between the peak-separation $\Delta^{\rm peak}$ 
and the photon energy $\epsilon$
may serve as a tool to verify some models of pulsar activity.

The presence of two peaks in gamma-ray pulses with large (0.4 - 0.5) phase separation may be understood
within a scenario of a single canonical polar cap (e.g. Daugherty \& Harding 1996, Miyazaki \& Takahara 1997).
One need to assume, however, a nearly aligned rotator, i.e. 
a rotator where three characteristic 
angles are of the same order: $\alpha$ - 
the angle  between spin axis $\vec \Omega$ and the  magnetic moment $\vec \mu$, 
$\theta_\gamma$ - the opening angle between a direction of 
the gamma-ray emission and $\vec \mu$,
and $\zeta$ - the angle between $\vec \Omega$ and the line of sight. 
For a canonical polar cap and instant electron acceleration, $\theta_\gamma$
roughly equals $0.02/\sqrt{P}$ radians only (where $P$ denotes a spin period).
To avoid uncomfortably small characteristic angles,
Daugherty \& Harding (1996) postulated
that primary electrons come from more extended polar caps,
and with the acceleration occuring at a height $h$ of several neutron-star radii $R_{\rm ns}$.
The latter assumption may be justified by a GR effect found by
Muslimov \& Tsygan (1992).

The aim of this paper
is to present general properties of $\Delta^{\rm peak}(\epsilon)$
obtained numerically (and to some degree semi-analytically) 
for four simplified versions of a polar-cap activity model.
In Section~2 we outline the model. Section~3
describes the results and offers some explanation of the presented effects. Conclusions follow in Section~4.

\section{The Model}

We use a polar cap model  with beam particles (primary electrons)
distributed evenly along a hollow cone formed by the magnetic field lines from 
the outer rim of a canonical polar cap,
i.e. with an opening angle $\theta_{\rm init} = \theta_{\rm pc}$, where
$\theta_{\rm pc} \simeq (2\pi \, R_{\rm ns}/c\,P)^{1/2}$ radians at
the stellar surface level ($h = 0$). In one case (model C - see below) we assume
$\theta_{\rm init} = 2 \theta_{\rm pc}$
The essential ingredients
of the high-energy processes 
had been introduced by Daugherty \& Harding (1982), with high-energy radiation due to 
curvature and synchrotron processes (CR and SR, respectively) induced by primary electrons 
accelerated to
ultrarelativistic energies. 
The pulsar parameters are those of the Vela: the spin period $P = 0.0893\s$,
and the dipolar magnetic field at the polar cap $B_{\rm pc} \approx 10^{12}\G$.

Within the hollow-cone geometry 
we have considered three scenarios for electron acceleration:\hfill\break
model A - beam particles are
injected at a height $h_{\rm init} = 0$ with some initial
ultrarelativistic  energy $E_{\rm init}$ (the values of $E_{\rm init}$ are listed
in Table 1) and no subsequent acceleration; \hfill\break
model B - similar to model A but beam
particles are injected at a height $h_{\rm init} = 1\, R_{\rm ns}$;\hfill\break
model C - beam particles are injected at a height $h_{\rm init} = 2\, R_{\rm ns}$ 
with a low energy $E_{\rm init}$ and then  
accelerated by a longitudinal electric field ${\cal E}$ present over 
a characteristic scale height $\Delta h = 0.6\, R_{\rm ns}$, resulting in total 
potential drop 
$V_0$:
\begin{equation}
{\cal E} (h) = \cases{ V_0/\Delta h, \,\, {\rm for} \,\, h_{\rm init} \leq h \leq (h_{\rm init}+\Delta h)\cr
0, \,\,\,\,\,\,\,\,\,\,\,\,\,\,\,\,{\rm elsewhere.}\cr }
\end{equation}

For comparison, we considered a model with 
a uniform electron distribution over 
the entire polar cap surface (i.e. $\theta_{\rm init} \in [\, 0, \theta_{\rm pc}]$):\hfill\break
model D - beam particles are
injected at a height $h_{\rm init} = 0$ with initial
ultrarelativistic  energy $E_{\rm init}$ as in the model A, and no subsequent acceleration. \hfill\break

The values of $E_{\rm init}$ in models A, B and D, and the potential drop
$V_0$ in model C were chosen to yield similar number of secondary pairs - about $10^3$ per beam particle.
Table 1. summarizes the properties of the models.

\def  \sep    #1{\raise8pt\hbox{#1} }
\def  \sepp   #1#2{ \vbox to 8.5mm { \hbox{#1} \vfill \hbox{#2} } }
\def  \above  #1#2{$\begin{array}{c} #1 \\
                                     \noalign{\vskip 5pt}
                                     #2 \end{array}$}
\def  \nolin       {\noalign{\vskip 2pt} \hline \noalign{\vskip 2pt}}

\footnotesize{
\vbox{
\begin{center}
Table 1. Model parameters.
\vskip 2mm
\begin{tabular}{c c c c c c}
\hline 
\noalign{\vskip 5pt}
    & \above{B_{\rm pc}}{[10^{12} \rm G]} & \above{\alpha}{[\rm deg]} 
       & \above{h_{\rm init}}{[R_{\rm ns}]} 
       & \above{\theta_{\rm init}}{[\theta_{\rm pc}]} & primary electrons \\


\noalign{\vskip 3pt}
\hline
\noalign{\vskip 4pt}
Model A & 1.0 & 3.0 & 0.0 & 1.0 & $E_{\rm init}=8.68$ TeV, no acceleration\\
\nolin
Model B & 1.0 & 5.0 & 1.0 & 1.0 & $E_{\rm init}=20.0$ TeV, no acceleration\\
\nolin
Model C & 3.0 & 10.0 & 2.0 & 2.0 &
$E_{\rm init}=0.5\MeV$, acceleration (see eq.(1))\\ 
& & & & & with $V_0 = 2.5\times 10^7$volts\\
\nolin
Model D & 1.0 & 3.0 & 0.0 & $[0,1]$ & 
$E_{\rm init}=8.68$\ \rm TeV,\ no\ acceleration\\
\nolin
\hline
\end{tabular}
\end{center}
}
}
\normalsize

\section{Results}

We have calculated  numerically the pulse shapes as a function of
photon energy for all four models.
The main difference between models A and B is due to different values of $h_{\rm init}$ which
result in different locations of origin of secondary particles.
Changing these locations is an easy way to modify spectral
properties of emergent radiation and enables to change (preferrably - to increase) the angle $\alpha$
as constrained by the observed energy-averaged peak
separation $\Delta^{\rm peak} \approx 0.43$ (Kanbach 1999).
For a given  model there are two possible
values of the angle $\zeta$ resulting in a desired peak 
separation $\Delta_{\rm peak}$ at some energy $\epsilon$ (which we
choose to be $\sim 1\GeV$), and we took the larger one in each case:
$\zeta = 3.75, 4.5, 15.$, and $3.65$ degrees for models A, B, C, and D, respectively.

In Fig.\t 1 we present dependency of peak separation on photon energy for all four models. 
Within a `low-energy` range (below a few GeV)  the peak separation either remains constant (Model B) or
slightly decreases with the increasing photon energy (Models A, C, and D).
At a critical energy \et~ around a few GeV, 
the separation \del\ undergoes a sudden turn: for $\epsilon > \epsilon_{\rm turn}$
it either rapidly increases (models A, B and C) or rapidly decreases (model D).
For comparison,
the overall trend (taking into account a substantial scatter of points) resulting 
from the analysis by Kanbach (see Fig.2, middle 
panel, of Kanbach 1999)
is marked schematically.

{\footnotesize
\begin{figure}

 \vspace*{-1cm}{
 \hspace*{-0.2cm}{
 \mbox{
 \hbox{
 \epsfxsize=17cm
 \epsfysize=8.5cm
 \epsffile{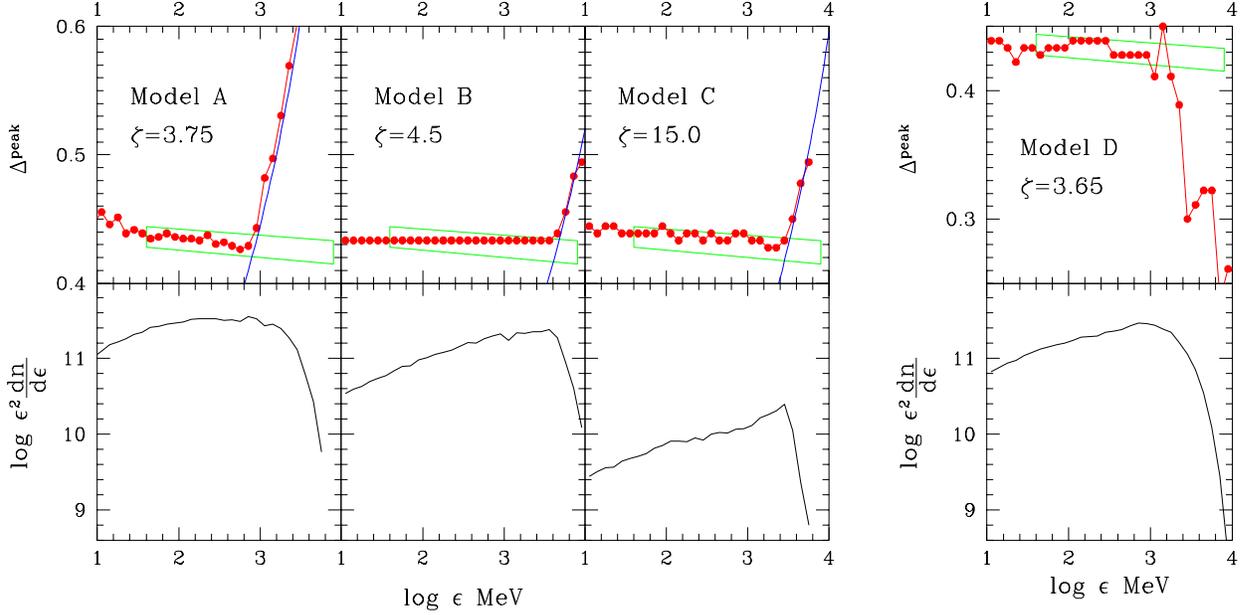}
 }}}}
\hskip -1cm
{\hspace{ 0cm}
\caption[]{Upper panels:
Peak separation \del\ against photon energy $\epsilon$, as predicted by 
models A, B, C and D (red dots). 
Steep blue lines in the three left panels (hollow-cone models) are due to semi-analytical
calculations for $\Delta^{\rm peak}(\epsilon)$ valid for $\epsilon > \epsilon_{\rm turn}$
(see text).
The overall observational trend, taking into account a substantial scatter of points (see Fig.2, middle 
panel, of Kanbach 1999),
is marked schematically for reference in all four panels
as a green parallelogram. The assumed values of $\zeta$ - the angle between $\vec \Omega$ and the 
line of sight - are also indicated.

Lower panels:
Energy output per logarithmic energy bandwidth 
at the first peak as a function of photon energy $\epsilon$
for models A, B, C and D.
[Note: these are not instantaneous spectra.]
Units on the
vertical axis are arbitrary.
}
\label{}
}
\end{figure}
}

It would be easy to understand why $\Delta^{\rm peak}$ should stay at a constant level
in the range $\epsilon < \epsilon_{\rm turn}$
if the gamma-ray photons were due exclusively to CR (especially in models with no acceleration).
A preferred direction of this emission would be set up already at the lowest magnetospheric altitudes.
This is because a monotonic decrease of the  electron's energy $E$,
and initial increase in the dipolar curvature radius 
$\rho_{\rm cr}$ make the level of contribution to the one-particle CR-spectrum at all energies
to be the highest one just at the initial altitude $h_{\rm init}$.
However, in the energy range below a few GeV the gamma-ray emission in our models is dominated by
the synchrotron radiation (SR) due to secondary $e^\pm$ pairs 
in the cascades. 
The resulting behaviour of $\Delta^{\rm peak}$ - either its slight decrease
with increasing $\epsilon$ (models A and C) or no change at all (model B) - 
is actually mediated by several factors which influence 
directional and spectral properties of the SR. These include
energy and pitch angle distributions of secondary $e^\pm$ pairs,
as well as the spatial spread of these pairs within the magnetospere (see e.g. Rudak \& Dyks 1999).
These factors do change from one model to another.

At the energy $\epsilon_{\rm turn}$ the separation $\Delta^{\rm peak}$ undergoes a sudden turn and starts
changing rapidly as $\epsilon$ increases. (This occurs above $\sim 1\GeV$, where most photons are
due to CR.)
This is the regime where the position of each peak in the pulse
is determined by the
magnetospheric opacity due to $\gamma \vec B \rightarrow e^\pm$. 

For the hollow-cone models (A, B and C)
the photons in both peaks of a pulse come from low magnetospheric altitudes with narrow opening angles. When 
$\epsilon$ is high enough these photons will be absorbed by the magnetic field with subsequent pair-creation.
In other words, inner parts of the `original' peaks in the pulse will be eaten-up and the gap between the peaks
(i.e. the peak separation) will increase. Photons which now found themselves in the `new' peaks come from
higher altitudes (the magnetosphere is transparent to them) and have wider opening angles.  
To quantify this line of arguments we present a simple semi-analytical solution for 
$\Delta^{\rm peak}$ as a function of $\epsilon$, which reproduces with astonishing accuracy
our Monte Carlo results (see Fig.1). 
For each point (at a given radial coordinate $r$) lying on
a magnetic field line  with a known angular coordinate $\theta$ on the stellar surface,
one defines the "escape energy" (Harding et al.\t 1997), which
is the upper limit \ee\ for photon energy if the photon is to
avoid magnetic absorption when propagating outwards.
We have approximated \ee\ with a power-law formula
$\epsilon_{\rm esc}(r) = a\left(r / R_{\rm ns}\right)^b$ MeV, and the values of $a$ and $b$
were found by fits to numerical solutions for each model.
We found 
$a\simeq 7.83\times 10^2$, $b\simeq 2.49$ for  $B=10^{12}\G$ and $\theta = \theta_{\rm pc}$ (models A and B);
$a\simeq 1.25\times 10^2$, $b\simeq 2.50$ for  $B=3\times 10^{12}\G$ and $\theta = 2\theta_{\rm pc}$ (model C).
Photons of some energy \eps\ will escape the magnetosphere only when they are 
emitted at $r\ge r_{\rm esc}$, where $r_{\rm esc}$ is the solution of the equation
$\epsilon_{\rm esc}(r)=\epsilon$. 
Let us now assume for simplicity, that photons which form the `new' peaks originate just at 
$r = r_{\rm esc}(\epsilon)$
(this is a reasonable assumption, especially for the models with no acceleration). 
With the coordinate $r$ of such an emitting ring determined, we now calculate the corresponding opening angle
(for dipolar magnetic field lines)
and then $\Delta^{\rm peak}$. 
In Fig.1 the blue lines represent the values of $\Delta^{\rm peak}(\epsilon)$ found semi-analytically,
while the red dots are the Monte Carlo results.
This branch of solution intersects the horizontal line
set by $\Delta^{\rm peak}=0.43$ at $\epsilon_{\rm turn}\simeq 0.9$, $4.5$, and $3$ GeV 
for models A, B, and C, respectively.

For model D, with uniform distribution of primary electrons over the polar cap (but otherwise identical
to model A) changes of \del\ above \et, occur in the opposite
sense. Unlike in previous models, here  both peaks of the pulse
are formed by
photons which were emitted along magnetic field lines attached to the polar cap
at some opening angle $\theta_{\rm init} < \theta_{\rm pc}$ . These photons are less 
attenuated than those coming from the 
outer rim, and in consequence the peak separation drops. Similar behaviour was obtained
by Miyazaki \& Takahara (1997) in their model of a homogeneous polar-cap.

Regardless the actual shape of the active part  (i.e. `covered' with primary electrons)
of the polar cap (either an outer rim,
or an entire cap, or a ring),
one does expect in general strong changes in the peak separation 
to occur at photon energies close to high-energy spectral cutoff due to 
magnetic absorption. 
To illustrate this, the lower panels of Fig.\t 1
present 
the energy output per logarithmic energy bandwidth 
at the first peak as a function of photon energy $\epsilon$
for models A, B, C and D.

\section{Summary}

Motivated by the recent suggestion (Kanbach 1999) that the peak separation
in gamma-ray pulses in the Vela pulsar may be energy dependent over the range
of $50\MeV$ to $9\GeV$, we calculated 
gamma-ray pulses expected in 
polar-cap models
with magnetospheric activity induced by curvature radiation of
beam particles.   Two types of geometry of 
a magnetospheric column above the polar cap were assumed: a hollow-cone column 
attached to the outer rim of the polar cap
and a filled column. Four models were considered with three scenarios for
beam acceleration.
The emission showing up as double-peak pulses is a superposition
of curvature radiation due to beam particles
and synchrotron radiation due to secondary $e^\pm$ pairs in cascades. 
The changes in the peak separation
were investigated with Monte Carlo numerical simulations and then reproduced (to some extent) with
semi-analytical methods.

We found that for the energy range $\epsilon < \epsilon_{\rm turn} \approx {\rm a \,\,few}\GeV$ 
the peak separation $\Delta^{\rm peak}$
either slightly decreases with increasing photon energy $\epsilon$ 
at the rate consistent with Kanbach (1999),
or stays at a constant level. The gamma-ray emission in this range
is dominated by synchrotron radiation in all four models considered.
The actual behaviour of $\Delta^{\rm peak}$ depends on  
physical properties of the pairs as well as on their spatial extent in the magnetosphere,
which vary from one model to the other.

At $\epsilon \simeq \epsilon_{\rm turn}$ the peak separation $\Delta^{\rm peak}$
makes an abrupt turn, and
for $\epsilon > \epsilon_{\rm turn}$
it changes dramatically. It increases in the hollow-cone models (A, B, C)
and decreases in the filled-column model (D), at a rate $\sim 0.28$ phase per decade
of photon energy. This is due to magnetic absorption effects ($\gamma \vec B \rightarrow e^\pm$).
The numerical behaviour of $\Delta^{\rm peak}$ in the hollow-cone models was reproduced with good accuracy 
with a simple 
semi-analytical approach to the condition of magnetospheric transparency for a photon
of energy $\epsilon$, originating at a given point in the dipolar magnetic field and
propagating outwards at a given direction.

The value of $\epsilon_{\rm turn}$ 
is model-dependent and for the cases considered here it stays 
between $\sim 0.9\GeV$ and $\sim 4.5\GeV$.   
To find such hypothetical turnover of $\Delta^{\rm peak}$
in real observational data
would require, however,
high-sensitivity detectors, since for $\epsilon > \epsilon_{\rm turn}$ the expected
flux of gamma-rays drops significantly (Fig.\t 1, lower panels).
If observed, this turnover would be a signature of polar cap activity in gamma-ray
pulsars, with high-energy cutoffs in their spectra due to magnetic absorption.
A detailed account of the effects presented above
will be given elsewhere (Dyks \& Rudak 1999, in preparation)

\section*{Acknowledgments}

This work has been financed by the KBN grants 2P03D-00911 and 2P03D-01016.
We are grateful to Gottfried Kanbach for discussions
on EGRET data of the Vela pulsar.

\section*{References}
\ni Daugherty, J.K., Harding, A.K., 1982, ApJ, 252, 337 \\
\ni Daugherty, J.K., Harding, A.K., 1996, ApJ, 458, 278\\
\ni Fierro, J.M., Michelson, P.F., Nolan, P.L., 1998, ApJ, 494, 734\\
\ni Harding A.K., Baring, M.G., Gonthier, P.L., 1997, ApJ, 476, 246\\
\ni Kanbach, G., 1999,  Proceedings of the 3rd INTEGRAL Workshop, in press\\
\ni Miyazaki, J., Takahara, F., 1997, MNRAS, 290, 49\\
\ni Muslimov,A.G., Tsygan, A.I., 1992, MNRAS, 255, 61\\
\ni Rudak, B., Dyks, J., 1999, MNRAS, 303, 477\\

\end{document}